# Stable magnetostructural coupling with tunable magnetoresponsive effects in hexagonal phase-transition ferromagnets

Enke Liu[1], Wenhong Wang[1], Lin Feng[1], Wei Zhu[1], Guijiang Li[1], Jinglan Chen[1], Hongwei Zhang[1], Guangheng Wu[1], Chengbao Jiang[2], Huibin Xu[2] and Frank de Boer[3]

The magnetostructural coupling between the structural and the magnetic transition plays a crucial role in magnetoresponsive effects in a martensitic-transition system. A combination of various magnetoresponsive effects based on this coupling may facilitate the multifunctional applications of a host material. Here, we demonstrate a possibility to obtain a stable magnetostructural coupling in a broad temperature window from 350 to 70 K, showing tunable magnetoresponsive effects, based on simultaneous manipulation of the phase stability and the magnetic structure by suitable chemical substitution of iron in MnNiGe. The resultant MnNiGe:Fe exhibits a magnetic-field-induced martensitic transition from paramagnetic austenite to ferromagnetic martensite, featuring (i) a large volume increase, (ii) a distinct magnetization change, (iii) small thermal hysteresis and (iv) a giant negative magnetocaloric effect. The results indicate that stable magnetostructural coupling is accessible in hexagonal phase-transition systems to attain the magnetoresponsive effects with broad tunability.

[1]State Key Laboratory for Magnetism, Beijing National Laboratory for Condensed Matter Physics, Institute of Physics, Chinese Academy of Sciences, Beijing 100190, China, [2]School of Materials Science and Engineering, Beihang University, Beijing 100083, China, [3]Van der Waals-Zeeman Instituut, Universiteit van Amsterdam, Amsterdam, Netherlands. Correspondence and requests for materials should be addressed to W.H.W. (e-mail: wenhong.wang@iphy.ac.cn) or F.R.deB. (e-mail: F.R.deBoer@uva.nl).

The ferromagnetic martensitic transition (FMMT)[1-3], a coinciding crystallographic and magnetic transition, is receiving increasing attention from both the magnetism and the material-science community due to the massive variations of associated magnetoresponsive effects, such as magnetic-field-induced shape memory (MFISM)/strain effect[4-7], magnetoresistance[8,9], Hall effect[10] and magnetocaloric effect (MCE)[11,12]. These effects are of interest for many potential technological applications like magnetic actuators[13,14], sensors[15], energy-harvesting devices[16] and solid-state magnetic refrigeration[17]. In these functionalities, the magnetostructural coupling between the structural and the magnetic transition plays an essential role. Seeking a stable coupling in a broad temperature range is a scientific and technological challenge.

In the case of ferromagnetic phase transitions coupled with martensitic-like structural changes, it is the ferromagnetic ordering (spontaneous magnetization) that triggers modest structural modifications due to the magnetoelastic coupling[18]. This magneto-elastic transitions have been utilized in the extensive and intensive investigations of a large body of giant magnetocaloric materials[19-27]. In contrast, in typical FMMTs, the change of structural symmetries of austenite and martensite is remarkable. The transformation thus converts the different magnetic states (moment values and type of coupling) in-between the two phases that have separate Curie (Néel) temperatures.

Since the MFISM effect based on the magnetostructural coupling has been discovered in the Ni-Mn-In Heusler alloy[7], attempts have been made to induce the structural transition of ferromagnetic shape memory alloys (FMSMAs) by applying a magnetic field. To this end, a large magnetization difference $\Delta M$ between the austenite and the martensite phase is important to maximize the magnetic-energy change introduced by applying a magnetic field. In a given system, if the martensitic transition is tuned to convert the magnetic states from the paramagnetic (PM) to the ferromagnetic (FM) state, rather than from FM to FM, a large $\Delta M$ will be found for the MFISM effect. Such a transition is scarcely observed in the case of ferromagnetic shape-memory alloys, including Fe-based alloys and Heusler alloys. With this transition, also a decrease of the magnetic entropy is associated. Therefore, it is of interest to find an alloy system which exhibits this particular magnetostructural transition, especially in a broad temperature range. For martensitic-transition systems, this PM-FM martensitic transition requires that the Curie temperature of the martensite should be higher than that of the austenite (see Fig. 1).

Recently, the MFISM effect has also been found in another type of materials[28], the hexagonal ternary compounds with $Ni_2In$ structure[29-31]. With the FMMTs in these materials, large MCEs are associated[32-35]. Very importantly, in these compounds the magnetic-ordering (Curie or Néel) temperatures of the martensite are higher than those of the austenite (see Fig. 1). This large material pool provides a platform for the expected magnetostructural transition mentioned above. As one of the important hexagonal materials, the stoichiometric MnNiGe compound undergoes a martensitic transition at a quite high temperature of $T_t$=470 K from the hexagonal ordered $Ni_2In$-type structure ($P6_3/mmc$, 194) to the orthorhombic TiNiSi-type structure ($Pnma$, 62) (refs 36, 37 and 38; see Fig. 1). Because this transition occurs in the PM state, the expected magnetostructural coupling cannot be established. Upon cooling, the martensite phase shows a magnetic transition from the PM state to the antiferromagnetic (AFM) state at a Néel temperature ($T_N^M$) of 346 K (ref. 36). The magnetic moments of 2.8 $\mu_B$, which are only localized on the Mn atoms, form an AFM spiral structure[36,37] so that the magnetization is very low. Besides, on the basis of data for near-stoichiometric MnNiGe systems (see detailed data in Supplementary Table S1), it can be estimated that the Curie temperature ($T_C^A$) of the high-temperature austenite of stoichiometric MnNiGe lies around 205 K, which is clearly below $T_N^M$ (346 K) of the low-temperature martensite. Thus, there is a large temperature interval of about 140 K between $T_C^A$ and $T_N^M$.

For modifying MnNiGe into a material which has the desired PM-FM martensitic transition, two important changes have to be introduced in the material. In the first place, the AFM transition in the martensite phase should be modified into a FM transition, i.e. $T_N^M$ should become $T_C^M$. This modification is indicated in Fig. 1 by the red dotted arrow and line. The second necessary modification is that the martensitic-transition temperature $T_t$ should be lowered in a controllable fashion to a temperature within the temperature interval, as indicated by blue dotted arrow in Fig. 1. Achievement of these two modifications would mean that a temperature window between $T_C^A$ and $T_C^M$ exists in which the magnetostructural coupling of the martensitic transition and the PM-FM transition is established with an appreciable value of $\Delta M$.

For realizing materials with this desired transition, it seems promising to substitute in MnNiGe the magnetic element Fe for the non-magnetic Ni or the magnetic Mn. This is promising because, in the isostructural compounds MnFeGe and FeNiGe, no martensitic transition occurs so that the austenite structure is maintained down to 4.2 K (see detailed data in Supplementary Fig. S1; ref. 29). At the same time, the magnetic Fe may alter the spirally AFM coupling of Mn moments in alloyed MnNiGe. In this sense, alloying these Fe-containing isostructures with MnNiGe may give rise to a more stable austenite (i.e. with lower $T_t$) and a higher magnetization of martensite (i.e. FM instead of AFM martensite). In the present investigation, we partly substituted Fe for Ni and Fe for Mn in MnNiGe, creating the quasi-ternary systems MnNi$_{1-x}$Fe$_x$Ge and Mn$_{1-x}$Fe$_x$NiGe.

## Results

**Sample preparation and characterization**. The samples were prepared by arc melting and homogenization annealing. The structure of samples was determined with X-ray diffraction (XRD) and no impurity phase was found. Details of the methods are given in the Methods section. With increasing Fe content, the transformation temperature from orthorhombic TiNiSi-type martensite to hexagonal Ni$_2$In-type austenite is gradually lowered from higher temperatures to below the room temperature (Fig. 2a). The XRD data show that the $c_h$ ($a_h$) axis of the austenite phase decreases (increases) upon Fe substitution (Fig. 2b). Temperature-dependent XRD reveals that the martensitic transition begins at 240 K in Mn$_{0.84}$Fe$_{0.16}$NiGe (Fig. 2c). An increase of 2.68% in unit-cell volume is found at the transition (Fig. 2d; see detailed data in Supplementary Table S2). This volume expansion is large and opposite to the usual contraction of about -1% at martensitic structural transitions. This indicates that the crystalline structure and the atomic surrounding undergo a pronounced reconstruction during the structural transition, as shown in the inset of Fig. 1.

**Structural and magnetic phase diagrams.** To determine the crystallographic and magnetic structures, low- and high-field $M(T)$ measurements and differential thermal analysis (DTA) were used (see detailed data in Supplementary Figs. S2, S3 and Table S3). Based on the experimental results, the MnNi$_{1-x}$Fe$_x$Ge and Mn$_{1-x}$Fe$_x$NiGe phase diagrams are proposed as shown in Fig. 3. In both systems, the Fe substitution makes the martensite ferromagnetic with a temperature of the martensitic transition that falls within the temperature range of ferromagnetic order.

Upon substitution of Fe for Ni (MnNi$_{1-x}$Fe$_x$Ge, see Fig. 3a), $T_N^M$ becomes $T_C^M$ at about 300 K. Upon further increase of Fe content, $T_t$ continuously decreases until $T_C^A$ is reached. It can also be seen that, upon substitution, both $T_N^M$ ($T_C^M$) and $T_C^A$ basically remain constant, which offers an accessible temperature window of about 90 K between $T_C^M$ and $T_C^A$. Within this window, the system undergoes a martensitic transition coupled with a magnetic transition from the PM to the FM state. Below $T_C^A$, the magnetostructural transition decouples as the martensitic transition rapidly vanishes. In the case of substitution of Fe for Mn (Mn$_{1-x}$Fe$_x$NiGe, see Fig. 3b), a quite low level of Fe substitution (about $x = 0.08$) already lowers $T_t$ to meet $T_N^M$ and to introduce ferromagnetism at a relatively high temperature of 350 K. In the range $0.08 \leq x \leq 0.26$, the FM martensite phase has a high magnetization at 5 T. The Fe substitution efficiently converts AFM martensite into FM martensite while, surprisingly, it drives the FM austenite parent phase into a weak-magnetic spin-glass-like state (see Fig. 4 and Supplementary Fig. S4). Along with the eventual vanishing of the martensitic transition at the freezing temperature ($T_g$) of the spin-glass-like state, the significant consequence is obtained: the lowest temperature of the window for the PM-FM martensitic transition with large $\Delta M$ has moved down to about 70 K and a quite broad temperature interval up to 280 K is generated for the stable magnetostructural coupling.

It should be pointed out that, the temperature hysteresis of the first-order martensitic transition for both systems is significantly reduced from about 50 K to below 10 K by the Fe substitution (see Supplementary Fig. S5 and Table S3), which implies a decreasing thermodynamic driving force for the martensite nucleation. For a first-order martensitic transition, the hysteresis of 10 K is very small[39] which is beneficial for the temperature sensibility of magnetoresponsive smart applications based on martensitic transitions.

**Thermomagnetic behavior.** To clarify the PM-FM transitions in the broad window, we measured the high-field thermomagnetic properties of the typical samples of $Mn_{1-x}Fe_xNiGe$ system, as shown in Fig. 4 (the thermomagnetic properties of $MnNi_{1-x}Fe_xGe$ system are shown in Supplementary Fig. S7). In accordance with Fig. 3, $T_t$ decreases with increasing Fe content. For x > 0.08, PM-FM jumps of the magnetization, with large $\Delta M$ up to 60 A m$^2$ kg$^{-1}$ in a field of 5 T, are observed. This signifies that the introduction of Fe has led to a great change of the magnetic exchange interaction in the martensite phase, changing the spiral AFM structure into a FM state. Upon cooling, for each composition the FM martensite phase nucleates and grows in the PM austenite matrix. Upon heating, the reversible nature of the martensitic transition can be seen. Here it should be emphasized again that $T_t$ is the martensitic-transition temperature, not the Curie temperature. The Curie (Néel) temperatures of both phases have the values at the respective window boundaries, which are shown in Fig. 3b. When the transition occurs, the austenite is still in PM state while the martensite is already in its FM state. It is the crystallographic structural transition between a PM phase and a FM phase that gives rise to the abrupt magnetization change. For $x = 0.26$, it can be seen that the martensitic transition becomes incomplete and the spin-glass-like behavior revealed by the irreversible ZFC-FC curves is in accordance with the phase diagram in Fig. 3b. Within the broad temperature window, a stable magnetostructural coupling is obtained from above room temperature (350 K) to liquid-nitrogen temperature (70 K).

**Magnetic-field-induced properties across the transitions.** In what follows, we study typical magnetoresponsive properties for both systems. Figure 5a shows the magnetization curves of $MnNi_{0.77}Fe_{0.23}Ge$ at various temperatures within the temperature window. Above 279 K, the austenite shows PM behavior. Between 274 and 258 K, the continuous metamagnetic behaviour at each temperature in high field (marked by arrows) reveals a distinct MFIMT effect, indicating that the FM martensite phase is induced by an applied field in the PM austenite matrix. This behavior corresponds to an upward shift of about 11 K of the martensitic starting transition temperature by a field of 5 T (see Supplementary Fig. S8), which means that the martensite phase appears at a higher temperature with the aid of the applied magnetic field. Based on the AFM-FM conversion of the martensite upon Fe substitution, the appreciable $\Delta M$ between the austenite and martensite introduces a larger Zeeman energy for the martensite in an applied magnetic field, giving rise to energetically more favourable martensite. Similarly, a more-distinct PM-FM MFIMT effect also occurs in $Mn_{0.82}Fe_{0.18}NiGe$, as shown in Fig. 5b. Based on this MFIMT effect, magnetic-field-controlled FMSMAs may be prepared in MnNiGe:Fe system. Moreover, this MFIMT with a large volume increase (see Fig. 2d) implies the volume of the material can be significantly changed by an applied field. This may benefit the magnetic-field-induced strains for potential strain-based applications.

Associated with the sharp first-order magnetostructural transition, a magnetic-entropy change ($\Delta S_m$) occurs[40,41]. By means of the Maxwell relation, the magnetic-entropy changes at the transitions have been derived from the magnetization curves of MnNi$_{1-x}$Fe$_x$Ge ($x = 0.23$) (Fig. 5c) and Mn$_{1-x}$Fe$_x$NiGe ($x = 0.18$) (Fig. 5d). Because the low-temperature martensite is FM and the high-temperature austenite is PM ($\partial M / \partial T < 0$), all samples exhibit a negative $\Delta S_m$. In the temperature window, MnNi$_{0.77}$Fe$_{0.23}$Ge exhibits a large $\Delta S_m$ value of -19 J kg$^{-1}$ K$^{-1}$ for $\Delta B = 0 - 5$ T (Fig. 5c). This window offers the possibility to obtain large $\Delta S_m$ values for the MnNi$_{1-x}$Fe$_x$Ge system in an interval of nearly 100 K. In the Mn$_{1-x}$Fe$_x$NiGe system, even more appreciable $\Delta S_m$ values are observed in an even more extended temperature window ranging from 350 to 70 K. As an example, a low substitution level of $x = 0.18$ gives rise to a giant $\Delta S_m$ value of -31 J kg$^{-1}$ K$^{-1}$ for $\Delta B = 0 - 5$ T (Fig. 5d). These larger $\Delta S_m$ values are attributed to the more-ferromagnetically ordered martensite and thus a lower magnetic-entropy state after the transition. In accordance with the MFIMT effect (see Supplementary Fig. S8), the $\Delta S_m$ peak position also shows field dependence and shifts to higher temperatures with increasing magnetic field (indicated by the dashed arrows in Figs. 5c and 5d). Another feature of the MnNiGe:Fe system is that the magnetic and martensitic transitions have the same sign of the enthalpy change (see Supplementary Figs. S2 and S3), since the crystallographic and magnetic symmetries are both lowered on cooling. This prevents opposite heat processes that counteract the caloric effects, which is very different from the common FM-AFM(PM) martensitic transitions in Fe-based and Heusler FMSMAs.

## Discussion

In this study, stable magnetostructural coupling has been realized by appropriate material design and the associated magnetoresponsive effects have been presented. This magnetostructural coupling has been achieved by decreasing $T_t$ of martensitic transition of alloyed MnNiGe and converting the AFM to the FM state in martensite phase by replacing Ni or Mn by Fe. We will discuss the possible origins of both achievements in this section.

As mentioned before, the isostructural MnFeGe and FeNiGe have a stable $Ni_2In$-type austenite structure without martensitic transition. This reasonably causes the decrease of the structural transition temperature $T_t$ (i.e., the increasingly stable austenite phase) upon substitution of Fe in MnNiGe, which is similar to the alloying effect in many alloys[42]. Inherently, this is related to the strengthening of local chemical bonds when Fe atoms are introduced at Ni or Mn sites. To get an insight into the change of chemical bonds, we have calculated the valence-electron localization function (ELF) for the $MnNi_{0.5}Fe_{0.5}Ge$ system (see Supplementary Fig. S9). The results indicate that substitution of Fe for Ni in MnNiGe leads to stronger covalent bonding between neighbouring Fe and Ge atoms and between neighbouring Mn and Mn atoms, which is thus largely responsible for the stabilization of the high-temperature austenite phase.

Next, we should address the origin of the conversion of the AFM to the FM state upon Fe substitution in the martensite. In the martensite phase of stoichiometric MnNiGe compound[36], the zero-moment Ni atoms are surrounded by six nearest-neighbour Mn atoms, forming a local Ni-6Mn configuration (see Fig. 6a, light blue zone; also see Supplementary

Fig. S1). The spiral AFM structure originates from the specific exchange interactions in the Mn moments in this specific moment and lattice configurations in the matrix. In high applied magnetic fields, this spiral AFM structure changes into a canted FM structure which eventually saturates ferromagnetically at about 10 T [ref. 36]. This behavior illustrates the instability of the spiral AFM magnetic structure in the stoichiometric MnNiGe martensite.

According to the atomic occupancy rule in ordered $Ni_2In$-type compounds (for more details, see information in Supplementary Fig. S1), Fe in $MnNi_{1-x}Fe_xGe$ will simply occupy the Ni sites. Similarly, Fe in $Mn_{1-x}Fe_xNiGe$ will occupy the Mn sites. Here we take $MnNi_{1-x}Fe_xGe$ system as an example. Due to the unchanged relative site occupation during the diffusionless and displacesive martensitic transition, all atoms consistently occupy their respective sites after the transition. In Fe-substituted martensite, whichever positions on Ni sites the Fe atoms occupy, some Ni-6Mn local atomic configurations change into Fe-6Mn ones. That is, crystallographically, every introduced Fe atom is always surrounded by six nearest-neighbour Mn atoms. With magnetic moments of $0.5 < \mu < 1$ $\mu_B$ (ref. 43) for Fe atoms, this Fe-6Mn local configuration internally establishes FM coupling. The spirally AFM-coupled Mn moments within the configuration is thus changed into parallel alignment due to the Fe moments. We schematically illustrate this FM Fe-centered local configuration (pink zones) in comparison with the spiral AFM matrix in Fig. 6a. This mechanism is similar to the FM exchange interaction established between Mn and Co atoms by substituting Co ($\mu = 1$ $\mu_B$) for Ni in MnNiGe (ref. 44). With increasing Fe content, the number of FM configurations will increase and they will overlap and form larger FM zones, in this way promoting the AFM-FM conversion in the martensite phase.

The above described FM coupling in local Fe-6Mn configurations is explicitly confirmed by the magnetization behaviour of MnNi$_{1-x}$Fe$_x$Ge martensites at 5 K in fields of up to 5 T, shown in Fig. 6b. The Fe-free sample shows AFM behaviour with a metamagnetic kink at a critical field $B_{cr}$ of 1.3 T, indicating a metamagnetic transition from a spiral AFM to a canted FM state, in accordance with the reported stoichiometric MnNiGe (ref. 36). In contrast, the Fe-substituted martensites show a large slope of the $M(B)$ curves in low fields that increases as a function of the Fe content. This suggests that an increasing FM component is generated in the system due to the existence of the Fe-centered Fe-6Mn configurations. Meanwhile, $B_{cr}$ rapidly decreases with increasing Fe content (Figs. 6b and 6d) which corresponds to an increasing ease for the applied field to change the spiral AFM structure to a forced parallel alignment. The larger the number of local Fe-6Mn configurations becomes, the lower $B_{cr}$ will be. For $x$ = 0.30 (the highest Fe content in the martensite phase due to the vanishing of the martensitic transition for higher substitution), a FM groundstate with a very low $B_{cr}$ of 0.05 T is found. Due to both the Fe substitution and the applied field, the magnetization reaches values of up to about 100 A m$^2$ kg$^{-1}$ in a field of 5 T, much higher than that for the Fe-free sample.

By substituting Fe for Mn in MnNiGe, the AFM-FM conversion is expedited (Fig. 6c). This is because the Fe atoms not only introduce FM coupling between Fe and Mn atoms, but also break up the consecutive AFM sublattices of the Mn moments. This rapidly makes the AFM order collapse. Thus, only a small Fe content of about $x$ = 0.08 is sufficient to get the maximal magnetization for Mn$_{1-x}$Fe$_x$NiGe (Fig. 6d). The samples more and more easily get magnetically saturated and show a rapidly decreasing saturation field (Figs. 6c and 6e).

The complete FM ground state appears in the sample with $x = 0.22$, versus $x = 0.30$ in MnNi$_{1-x}$Fe$_x$Ge (Fig. 6c). Therefore, we conclude that the AFM-FM conversion becomes more efficient when the substituted Fe atoms occupy Mn sites as occurs in Mn$_{1-x}$Fe$_x$NiGe. As a consequence, this FM state in martensite phase facilitates the magnetoresponsive effects presented in this study.

In previous studies[44,45], it has been reported that the large-size Ge and zero-moment Ni in MnNiGe can be replaced by small-size Si and magnetic Co, respectively. Actually, these replacements are also equal to alloying the isostructural Ni$_2$In-type MnNiSi and MnCoGe compound with the MnNiGe mother compound. MnNiSi and MnCoGe undergo martensitic transitions at high temperatures and their martensites are both ferromagnetic[38]. Therefore, in martensite structure, MnNiSi and MnCoGe can reasonably change the AFM state of alloyed MnNiGe into a FM state; however, they fail to lower the martensitic transition temperature from 470 K of MnNiGe to below $T_C^M$ to establish the coupling needed for the desired magnetostructural transition. The substitution of Fe applied in this study thus shows the best results for both the decrease of $T_t$ and the magnetic-state conversion of MnNiGe, that is, for the desired PM-FM magnetostructural transition. Very recently, an interesting paper has been published on the pressure-tuned magnetostructural transition in Cr-doped MnCoGe (ref. 46). This paper clearly clarifies the importance and tunability of magnetostructural coupling in these hexagonal ferromagnets.

To summarize, we have found that a stable PM-FM magnetostructural coupling in a broad temperature window, with tunable magnetoresponsive properties, can be obtained in martensitic-phase-transition materials tailored by suitable substitution of Fe in MnNiGe. It has been found that the MnNiGe can be manipulated in terms of crystallology and magnetism to be easily affected by an applied magnetic field. The Mn-based MnNiGe:Fe material has been shown to possess compelling thermodynamic, crystallographic and magnetoresponsive effects with broad tunability in a broad temperature range. These effects may be utilized in potential applications working in the range between room temperature and liquid nitrogen temperature, such as magnetic-field-controllable martensite-particle/substrate composites[3,47], solid-state magnetic refrigeration[12,24,27] or multifunctional phase-transition-strain/magnetic sensors[15] jointly driven by both large-strain structural transition and sensitive magnetic switching from the PM to the FM state. The presented scheme of designing materials with appropriate magnetostructural transition may be of importance in the explore of multifunctional magnetoresponsive materials among new and known magnetic martensitic-transition systems.

## Methods

**Sample preparation.** Polycrystalline ingots were prepared by arc melting high-purity metals in argon atmosphere. The ingots were melted four times and turned over in-between to guarantee good alloying. The ingots were subsequently homogenized by annealing in evacuated quartz tubes under argon at 1123 K for five days and slowly cooled at 1 K min[-1] to room temperature to avoid stress in samples.

**Structure and thermal analysis.** The samples for powder x-ray diffraction (XRD) were made by fine grinding. The room-temperature powder XRD measurements were performed using a Rigaku XRD D/max 2400 diffractometer with Cu-$K_\alpha$ radiation. Temperature-dependent XRD measurements were performed from 285 to 98 K with a cooling rate of 2.5 K min$^{-1}$ using a Bruker XRD D8-Advance diffractometer with Cu-$K_a$ radiation. At each temperature, a waiting time of 30 min was included before starting the measurement. Differential thermal analysis (DTA) with heating and cooling rates of 2.5 K min$^{-1}$ was used to determine the martensitic-transition characteristic temperatures (defined as the peak values of the dM/dT and DTA curves).

**Magnetic measurements.** Magnetization measurements were carried out on powder samples using a superconducting quantum interference device (SQUID, Quantum Design MPMS XL-7). Low field $M(T)$ measuments on high-temperature PM austenite and low-temperature martensite were performed to study the temperature-dependent magnetic behaviour and also to determine the martensitic-transition characteristic temperatures. These experiments were combined with DTA. ZFC-FC thermalmagnetization in a field of 0.01 T and frequency-dependent magnetic susceptibility were measured with frequencies $f$ = 1, 97, 496, 997 and 1488 Hz in an AC magnetic field of 4 Oe after ZFC from 300 K.

In order to accurately derive the magnetic-entropy changes ($\Delta S_\mathrm{m}$) at magnetostructural transitions with thermal hysteresis, the so-called *loop process* method[48] was adopted to get the isothermal magnetization curves. The isothermal $M(B)$ curves were measured in fields of up to 5 T upon cooling with a temperature interval of 2 K. Before each isothermal magnetization, the samples were all the way heated up to the complete PM austenite region (100 K away from the magnetostructural transitions) with heating rate of 5 K min$^{-1}$ to eliminate the history-dependent magnetic states and then cooled back to the targeted measurement temperature points. All these temperature loops were performed in zero field. To avoid the overmuch temperature-induced FM martensite phase during approaching each targeted temperature point, the temperature scanning mode was set as NO OVERSHOOT with cooling rate of 2 K min$^{-1}$. Besides, a waiting time of 300 s was compelled before the measurements to guarantee a highly stable temperature. The $\Delta S_{\mathrm{m}}$ was then derived from the resulting magnetization curves using the Maxwell relation[49]:

$$\Delta S_{\mathrm{m}}(T,H) = S_{\mathrm{m}}(T,H) - S_{\mathrm{m}}(T,0) = \int_0^H \left( \frac{\partial M}{\partial T} \right)_{\mathrm{H}} dH.$$

**ELF calculations.** The ELF calculations using the pseudopotential method with plane-wave-basis set based on density-functional theory were presented in Supplementary Methods.

# Acknowledgements

The authors gratefully acknowledge Dr. Yu Wang (Xi'an Jiaotong University, China) for kind help in temperature-dependent XRD measurements, Prof. Xiaobing Ren (NIMS, Japan) and Prof. V. A. Chernenko (Universidad del País Vasco, Spain) for fruitful discussions. This work was supported by the National Natural Science Foundation of China (51031004 and 51021061) and National Basic Research Program of China (2009CB929501).

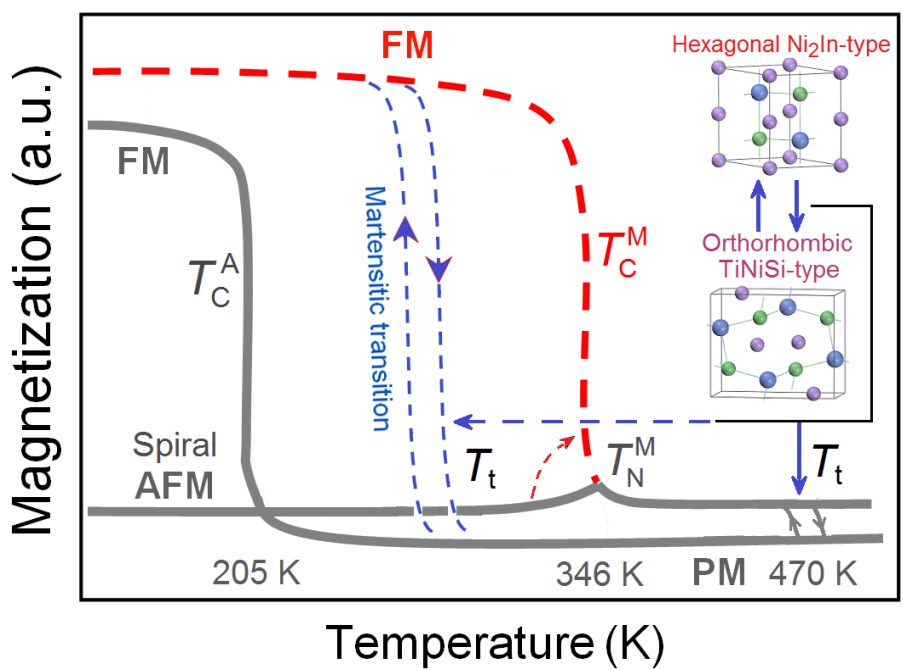

**Figure 1 | Expected magnetostructural coupling based on martensitic and PM-FM magnetic transitions.** The gray solid lines display the magnetization of the ferromagnetic Ni$_2$In-type austenite with Curie temperature ($T_C^A$) and the magnetization of the antiferromagnetic TiNiSi-type martensite with Néel temperature ($T_N^M$). At $T_t$ = 470 K, the hexagonal Ni$_2$In-type to orthorhombic TiNiSi-type martensitic transition occurs in stoichiometric MnNiGe, indicated by the blue solid arrow. The red dashed arrow indicates the expected AFM-to-FM conversion and the blue dashed arrow the expected decrease of $T_t$. Within the expected temperature window between $T_C^A$ and $T_C^M$, the martensitic transition would couple with a magnetic transition from the PM to the FM state.

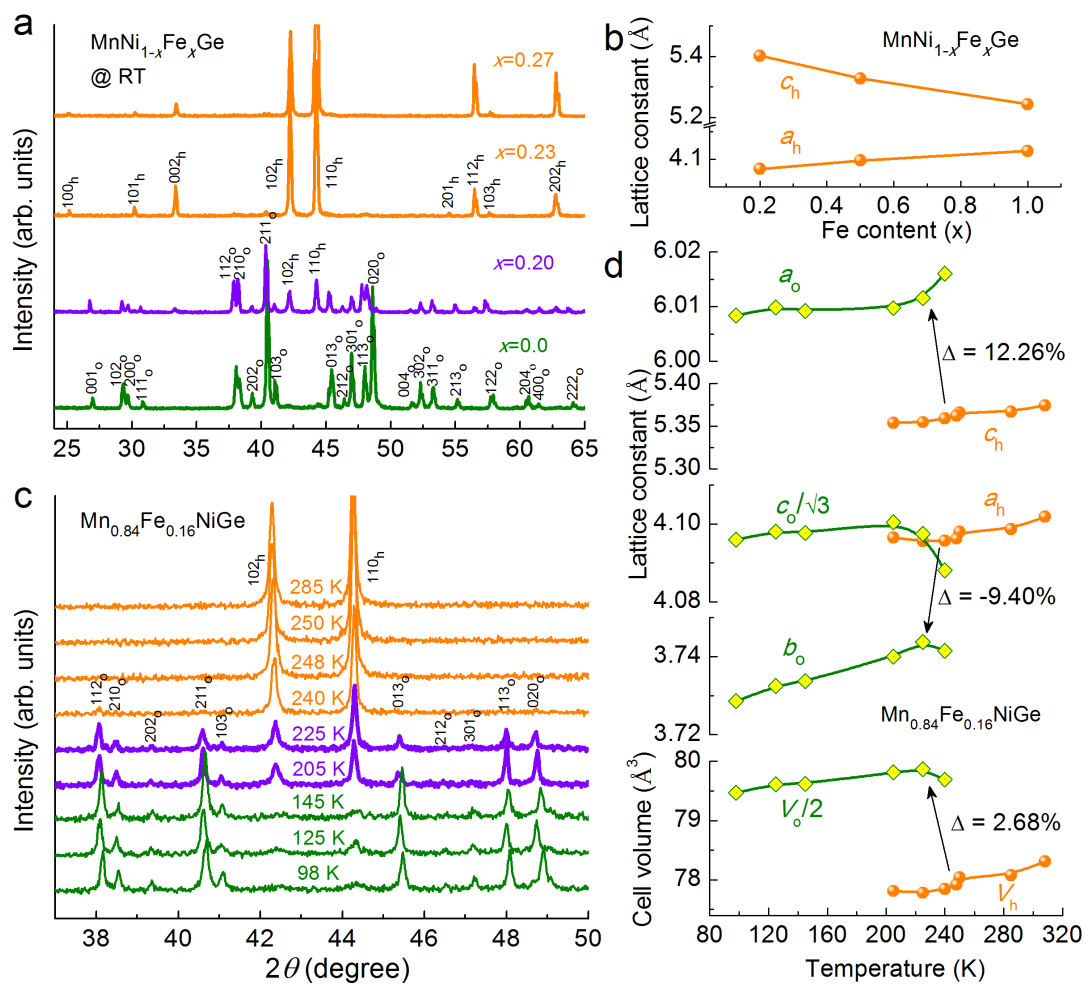

**Figure 2 | XRD analysis of the phase structures and phase transitions.** $hkl_h$ and $hkl_o$ denote the Miller indices for the hexagonal and the orthorhombic structure, respectively. (**a**), XRD of MnNi$_{1-x}$Fe$_x$Ge at room temperature. (**b**), Temperature-dependent XRD of Mn$_{0.84}$Fe$_{0.16}$NiGe from 285 to 98 K, indicating the martensitic transition from hexagonal to orthorhombic structure. The dashed lines in (**a**) and (**b**) denote the Bragg reflections of the hexagonal austenite phase. The remaining reflections are of the martensite phase. (**c**), Variation of the $a_h$ ($c_h$) axis of the austenite phase with Fe content. (**d**), Temperature dependence of the lattice constant and the cell volume of Mn$_{0.84}$Fe$_{0.16}$NiGe across the martensitic transition. The axes and volumes of the two structures are related as $a_o = c_h$, $b_o = a_h$, $c_o = \sqrt{3}a_h$ and $V_o = 2V_h$ (ref. 38).

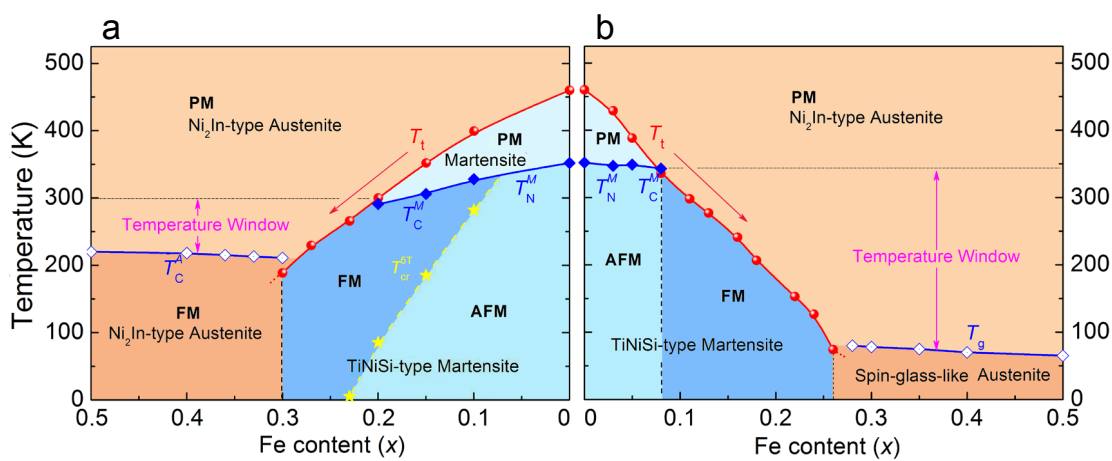

**Figure 3 | Structural and magnetic phase diagrams.** The red solid circles denote the

martensitic-transition temperature $T_t$ and the red arrows indicate the decreasing trend of $T_t$. In the range

$0 \leq x \leq 0.30$ for MnNi$_{1-x}$Fe$_x$Ge (**a**) and $0 \leq x \leq 0.26$ for Mn$_{1-x}$Fe$_x$NiGe (**b**), the systems undergo a Ni$_2$In-type to TiNiSi-type martensitic transition at $T_t$. Above $x = 0.30$ (**a**) and $x = 0.26$ (**b**), the systems are single-phase Ni$_2$In-type austenites. The blue solid diamonds correspond to $T_N^M$ ($T_C^M$) of the martensite

(**a, b**) and the blue open diamonds to $T_C^A$ (**a**) and $T_g$ (**b**) of the austenite. In the composition range

$0.20 \leq x \leq 0.30$ (**a**) and $0.08 \leq x \leq 0.26$ (**b**), the temperature windows are limited by $T_C^M$-$T_C^A$ (**a**) and by

$T_C^M$-$T_g$ (**b**), respectively. In **a**, the FM state of the martensite returns back to the AFM state (see

Supplementary Fig. S6). The yellow-star line corresponds to the critical temperature ($T_{cr}^{5T}$) between the

FM and AFM states in a field of 5 T. In **b**, the austenite-phase zone with $x \geq 0.26$ enters into a spin-glass-like state below about 70 K (see Supplementary Fig. S4).

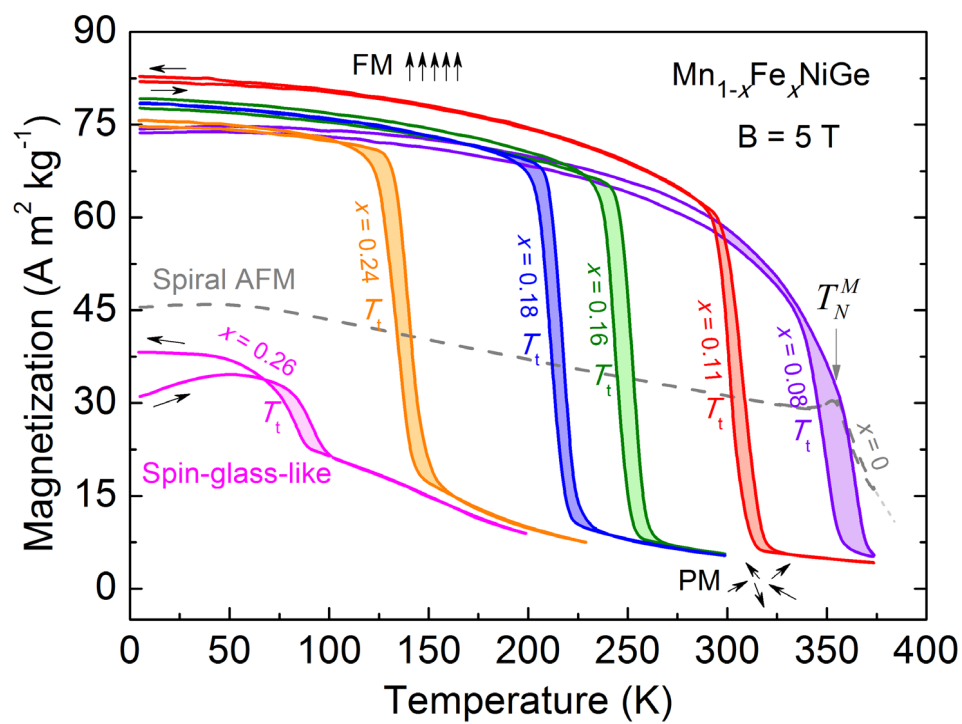

**Figure 4 | Stable magnetostructural coupling in the broad temperature window in Mn$_{1-x}$Fe$_x$NiGe system.** The magnetization curves were measured in the zero-field-cooling and field-cooling modes in an applied magnetic field of 5 T. $T_t$ indicates the martensitic-transition temperature. "PM" and "FM" indicate the PM austenite and FM martensite phases, respectively. The dashed curve represents the spiral AFM martensite of Fe-free stoichiometric MnNiGe. For stoichiometric MnNiGe, $T_N^M$ in high field is 350 K and $T_t$ is 460 K. An incomplete martensitic transition and spin-glass-like behavior are observed for the compound with x = 0.26 (see Supplementary Fig. S4).

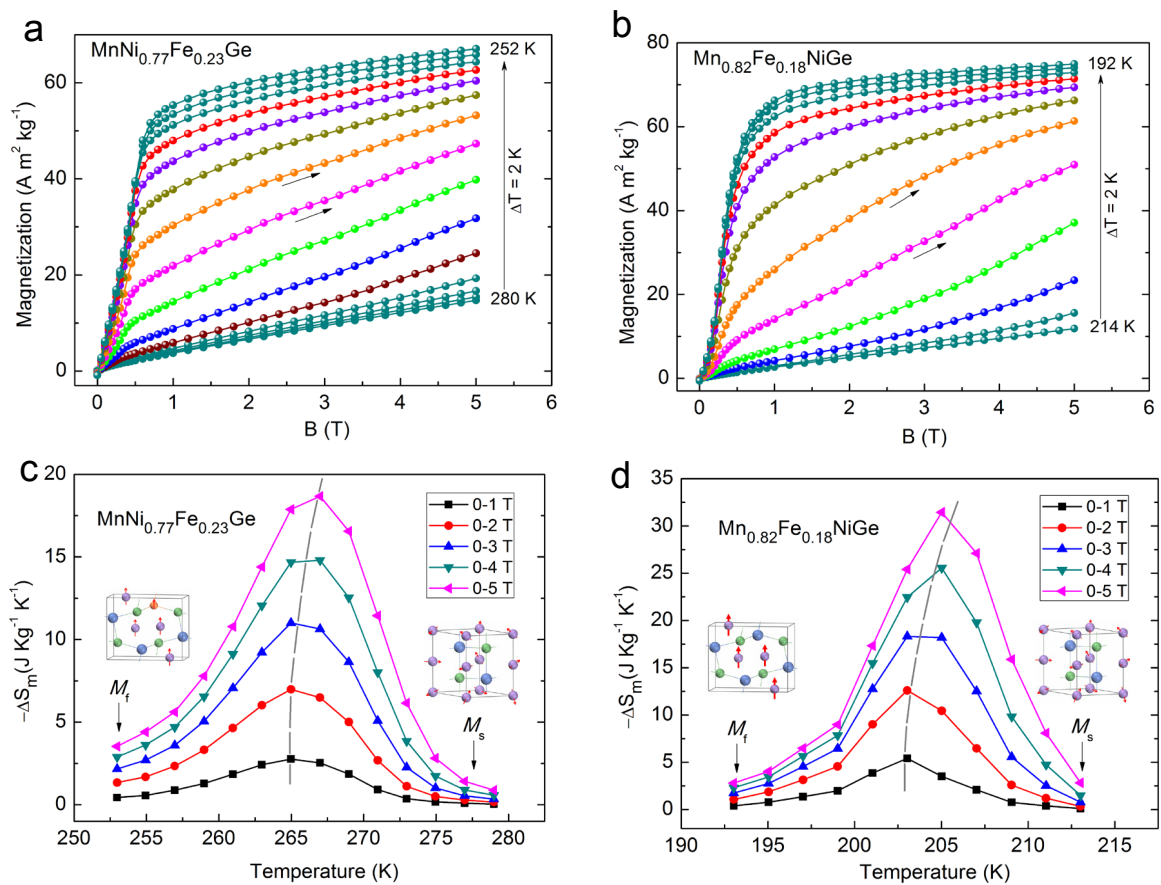

**Figure 5 | Magnetic-field-induced martensitic transitions and associated magnetic-entropy changes.** Magnetic isotherms of MnNi$_{0.77}$Fe$_{0.23}$Ge (**a**) and Mn$_{0.82}$Fe$_{0.18}$NiGe (**b**) at various temperatures in the temperature window. The metamagnetic behavior indicates the magnetic-field-induced martensitic structural transition. Isothermal magnetic-entropy changes ($\Delta S_m$) for various field changes derived from the isothermal magnetization curves of MnNi$_{1-x}$Fe$_x$Ge (**c**) and Mn$_{1-x}$Fe$_x$NiGe (**d**). The PM state in austenite and the FM state in martensite are depicted in the unit cells as insets in (**c**) and (**d**). $M_s$ and $M_f$ denote the starting and finishing temperatures of the martensitic transition, respectively.

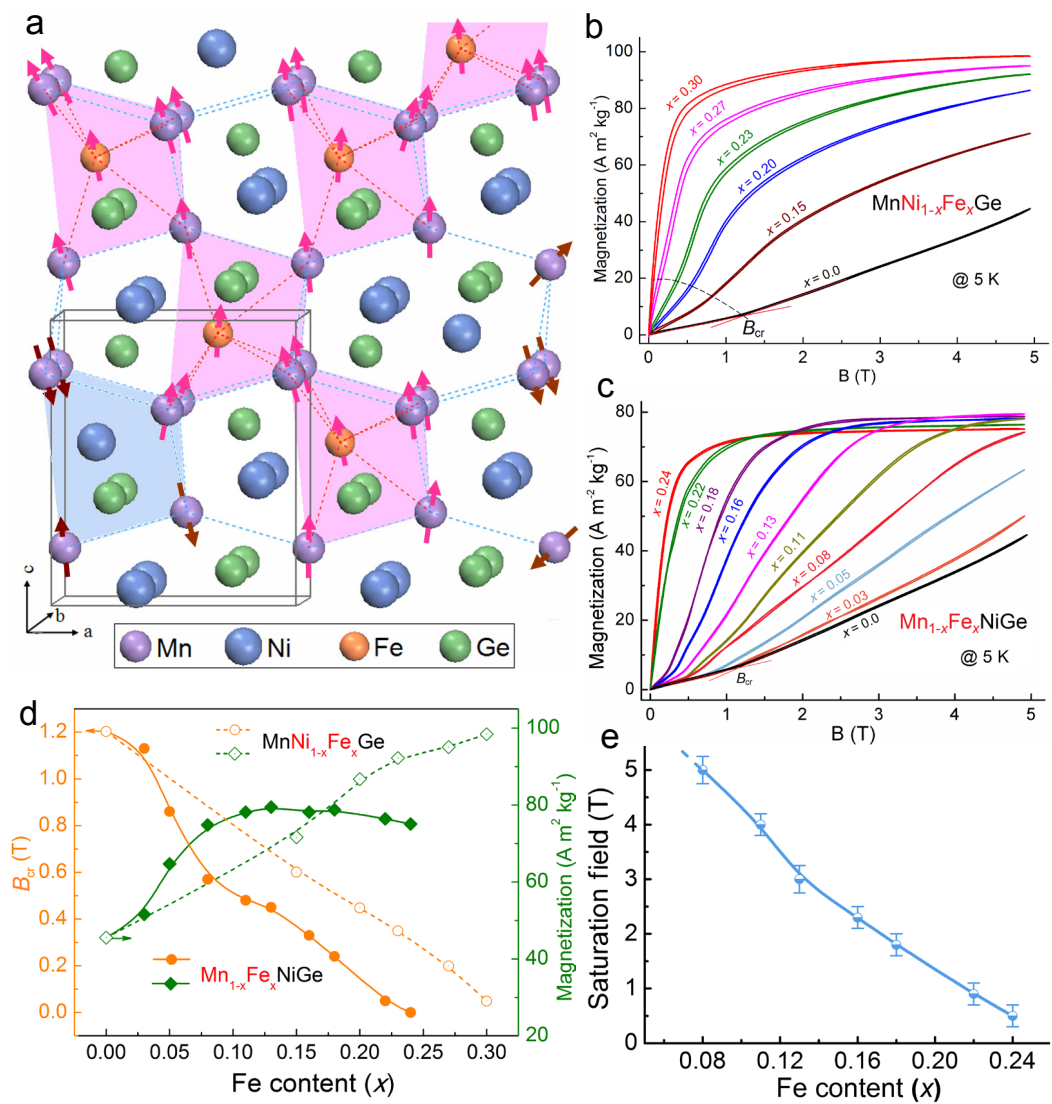

**Figure 6 | AFM-FM conversion in MnNiGe:Fe martensite. a**, Crystal structure of TiNiSi-type (*Pnma*, 62) MnNi$_{1-x}$Fe$_x$Ge martensite with indicated Ni-6Mn local configurations (light blue zones) and FM Fe-6Mn local configurations (pink zones). All atoms are at 4c (*x*, 1/4, *z*) positions[36,37]. The Fe atoms occupy Ni sites[29] and each Ni (Fe) atom is surrounded by six nearest-neighbour Mn atoms. **b**, **c**, Experimental evidence of (spiral AFM)-FM conversion as a function of both Fe content (*x*) and applied field (*B*) for both alloying cases. The *M*(*B*) curves have been measured at 5 K at which the samples are in the martensite state. The kink at $B_{cr}$ corresponds to the metamagnetic critical field of martensite. The dashed line indicates the decrease of $B_{cr}$. **d**, Dependence of $B_{cr}$ and the magnetization at 5 T of

MnNi$_{1-x}$Fe$_x$Ge martensite on the Fe content. **e**, Dependence of the saturation field of Mn$_{1-x}$Fe$_x$NiGe martensite on the Fe content.

# Stable magnetostructural coupling with tunable magnetoresponsive effects in hexagonal phase-transition ferromagnets

Enke Liu[1], Wenhong Wang[1], Lin Feng[1], Wei Zhu[1], Guijiang Li[1], Jinglan Chen[1], Hongwei Zhang[1], Guangheng Wu[1], Chengbao Jiang[2], Huibin Xu[2] and Frank de Boer[3]

1. *Beijing National Laboratory for Condensed Matter Physics, Institute of Physics, Chinese Academy of Sciences, Beijing 100190, China*

2. *School of Materials Science and Engineering, Beihang University, Beijing 100083, China*

3. *Van der Waals-Zeeman Instituut, Universiteit van Amsterdam, Amsterdam, Netherlands*

# SUPPLEMENTARY INFORMATION

a

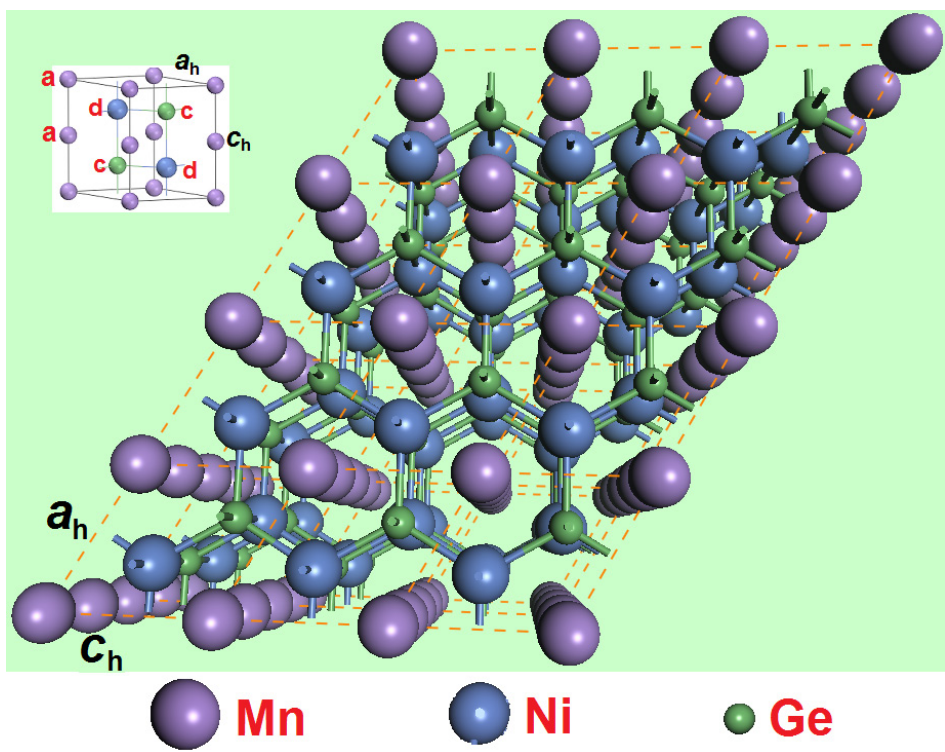

Mn     Ni     Ge

b

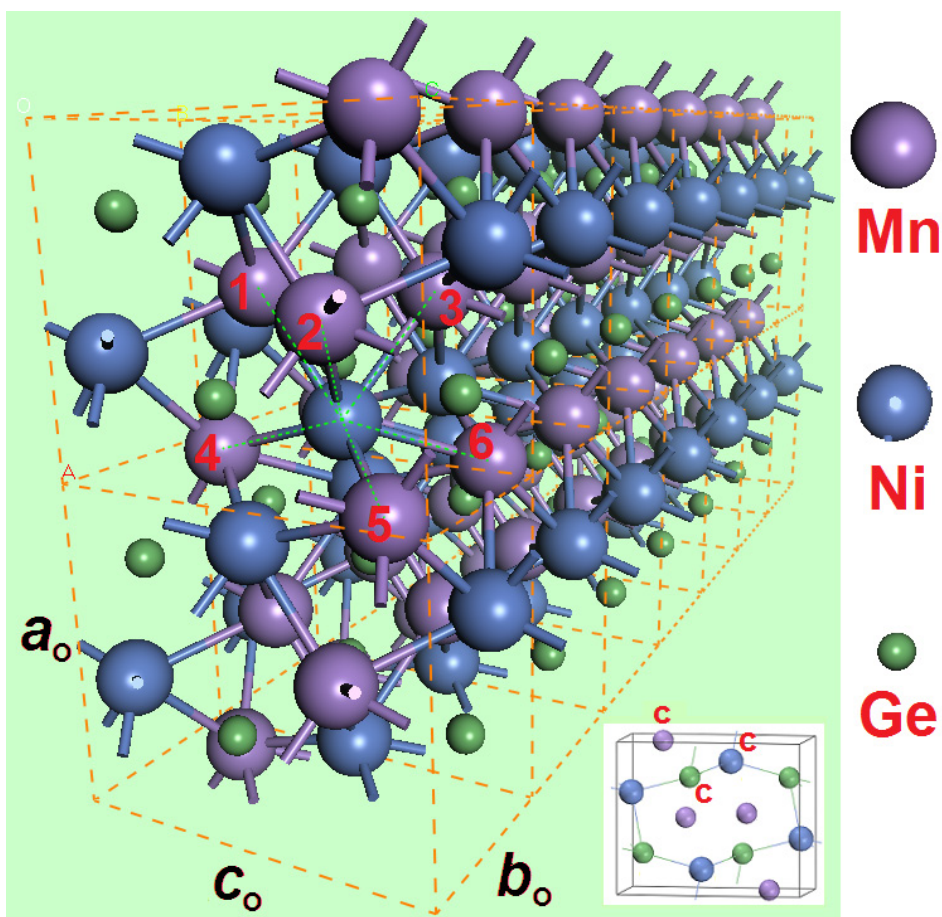

Mn

Ni

Ge

# SUPPLEMENTARY INFORMATION

## Supplementary Figure S1:

**Crystal structures of Ni$_2$In-type (*P*6$_3$/*mmc*, 194) (a) and TiNiSi-type (*Pnma*, 62) (b) of MnNiGe**. **a**, Layered structure of hexagonal austenite phase. The atomic occupancy sites in MnNiGe are Mn: 2a sites, Ni: 2d sites, Ge: 2c sites[1]. The inset is the unit cell of Ni$_2$In-type structure. **b**, Distorted layered structure of orthorhombic martensite phase. All the atoms are in 4c (x, 1/4, z) sites[1,2]. The inset is the unit cell of TiNiSi-type structure.

**Axe and volume relations of two structures**. The unit cell axes and volumes of the two structures are related as $a_o=c_h$, $b_o=a_h$, $c_o=\sqrt{3}a_h$ and $V_o=2V_h$ (ref. 3).

**Isostructures**. As Ni$_2$In-type ternary compounds[1,4], MnNiGe, MnFeGe and FeNiGe are all isostructural ordered intermetallics. The neutron diffraction experiments have determined that their space group number is 194 and the atomic occupancy sites are as follows (see inset of Figure S1**a**):

   Mn: 2a sites, Ni: 2d sites, Ge: 2c sites in MnNiGe;

   Mn: 2a sites, Fe: 2d sites, Ge: 2c sites in MnFeGe;

   Fe: 2a sites, Ni: 2d sites, Ge: 2c sites in FeNiGe.

This atomic occupancy rule is dominated by covalent electron numbers[3]. In a given Ni$_2$In-type system, the element with more covalent electrons preferentially occupies 2a sites, while the element with less covalent electrons preferentially occupies 2d sites. The *p*-block elements always occupy 2c sites. That is, the Ni atoms in Ni$_2$In-type MnNiGe and Fe atoms in isostructural MnFeGe both occupy the same 2d sites. Therefore, in MnNi$_{1-x}$Fe$_x$Ge the Fe will simply occupy the Ni (2d) sites. Similarly,

# SUPPLEMENTARY INFORMATION

alloying isostructural FeNiGe with MnNiGe ($Mn_{1-x}Fe_xNiGe$), the Fe will simply occupy the Mn (2a) sublattices.

**Ni-6Mn and Fe-6Mn local configurations**. In TiNiSi-type martensite phase of stoichiometric MnNiGe (refs 1, 2), each zero-moment Ni atom is surrounded by six nearest-neighbour Mn atoms, forming a Ni-6Mn local atomic configuration (see Figure S1**b**, indicated by green dashed lines). After some Ni atoms are replaced by Fe atoms, the Fe atoms are surrounded by six nearest-neighbour Mn atoms. That is, the Fe-6Mn local atomic configurations are formed in the same positions.

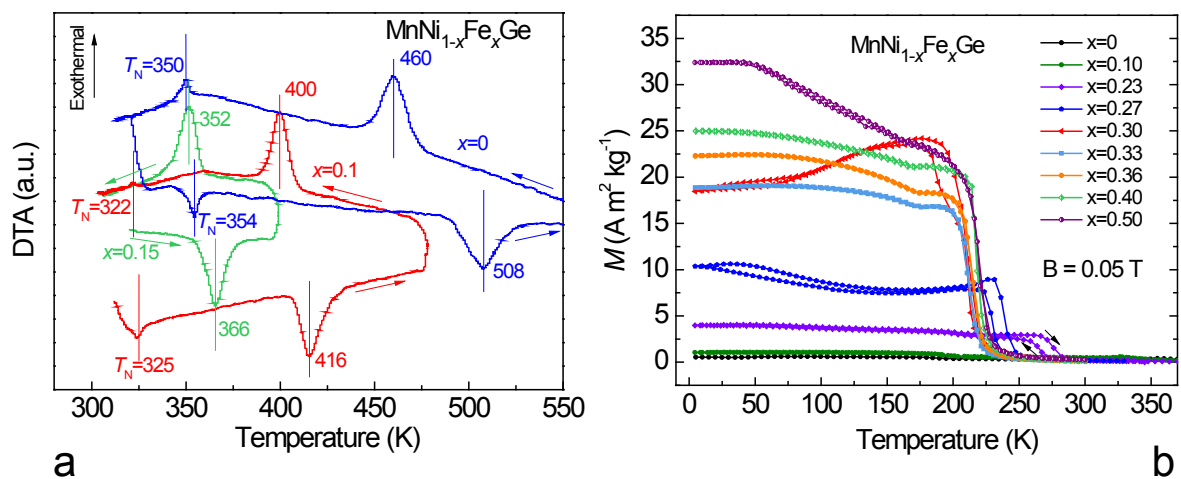

## Supplementary Figure S2:

**Determination of the magnetic and martensitic transitions by DTA, low-field M(T) curves and XRD for MnNi$_{1-x}$Fe$_x$Ge. a,** For samples with their martensitic transition occurring in the PM state, DTA with heating and cooling rates of 2.5 K/min was used to detect the transition data. $T_t$s were determined as the DTA peak values. $T_N$ of martensite phase on cooling and heating were indicated in the figure. The magnetic and martensitic transitions have the same sign of the enthalpy change upon heating and cooling. **b,** For samples with their martensitic transition occurring in the AFM (FM) state, M(T) curves were measured in a field of 0.05 T. $T_t$ was determined as the temperature where dM/dT has its maximum at the martensitic transition.

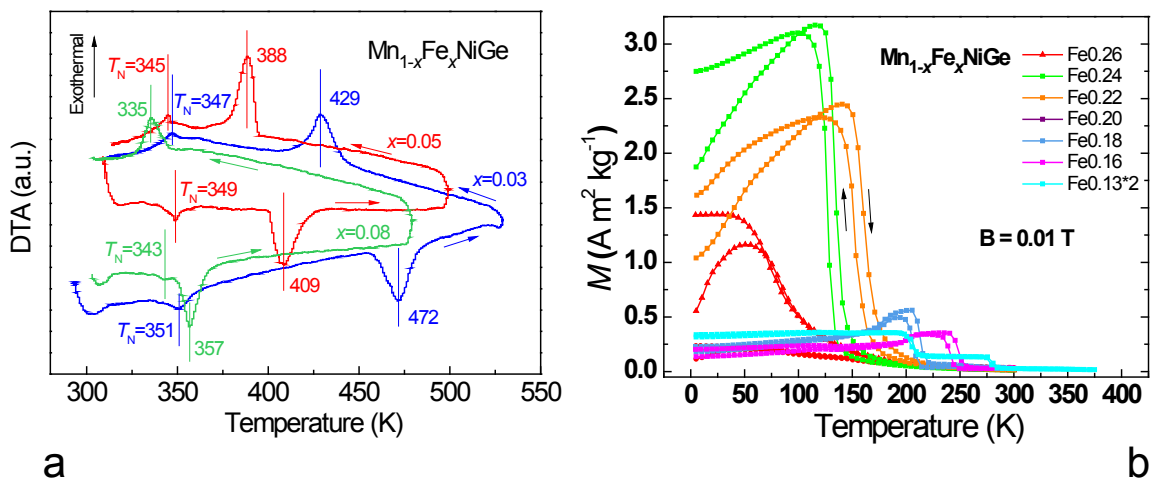

## Supplementary Figure S3:

**Determination of the magnetic and martensitic transitions by DTA, low-field M(T) curves for Mn$_{1-x}$Fe$_x$NiGe**. **a**, For samples with their martensitic transition occurring in the PM state, DTA with heating and cooling rates of 2.5 K/min was used to detect the transition data. $T_t$ were determined as the DTA peak values. $T_N$ of martensite phase on cooling and heating were indicated in the figure. The magnetic and martensitic transitions have the same sign of the enthalpy change upon heating and cooling. **b**, For samples with their martensitic transition occurring in the AFM (FM) state, M(T) curves were measured in a field of 0.01 T. $T_t$ was determined as the temperature where dM/dT has its maximum at the martensitic transition.

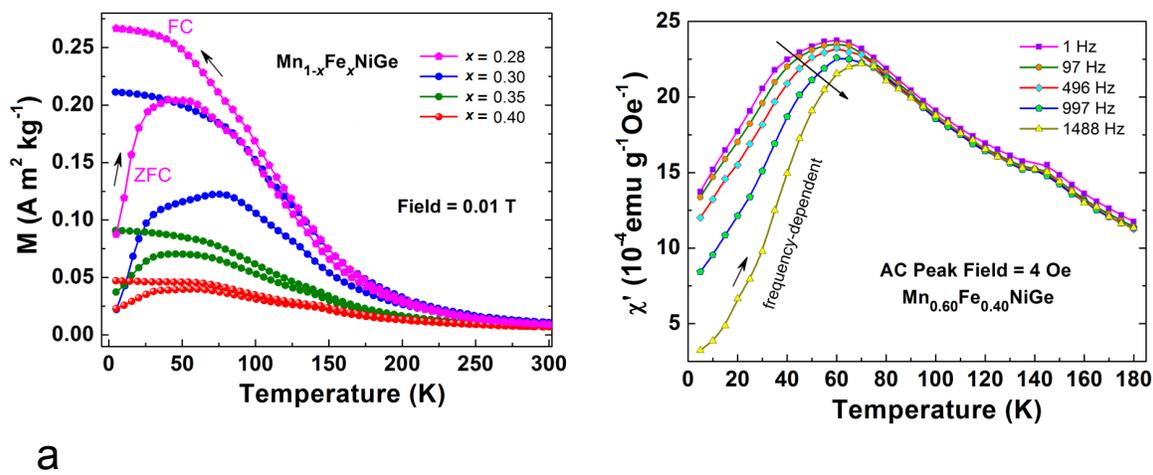

a                                                                                                                    b

## Supplementary Figure S4:

**Spin-glass-like behaviors of the austenite with *x* > 0.26. a,** Irreversible ZFC-FC curves of samples whit *x* = 0.28, 0.30, 0.35 and 0.40 in a low field of 0.01 T. The magnetization curves show weak magnetic spin-glass-like austenite with frozen temperature ($T_g$) at about 70 K. It may be understandable as a consequence of alloying PM FeNiGe with FM MnNiGe. **b,** Temperature dependence of the real part of the AC susceptibility, $\chi'$, of sample with *x* = 0.40 measured at frequencies f = 1, 97, 496, 997 and 1488 Hz in an AC magnetic field of 4 Oe after ZFC from 300 K. A clear frequency-dependent peak position ($T_g$ shifting to high temperatures) further indicates the spin-glass-like behaviors.

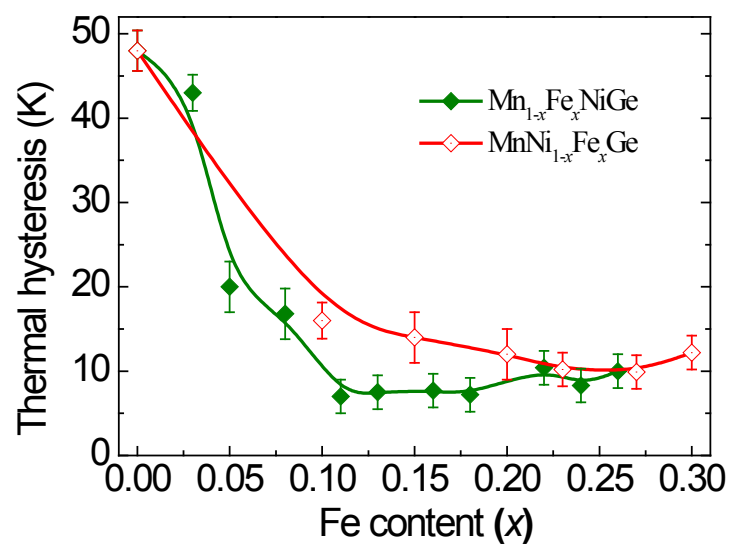

## Supplementary Figure S5:

**Temperature hysteresis of the first-order martensitic transition for Mn$_{1-x}$Fe$_x$NiGe and MnNi$_{1-x}$Fe$_x$Ge.** The phase-transition hysteresis is significantly reduced from about 50 K to below 10 K by the Fe substitution.

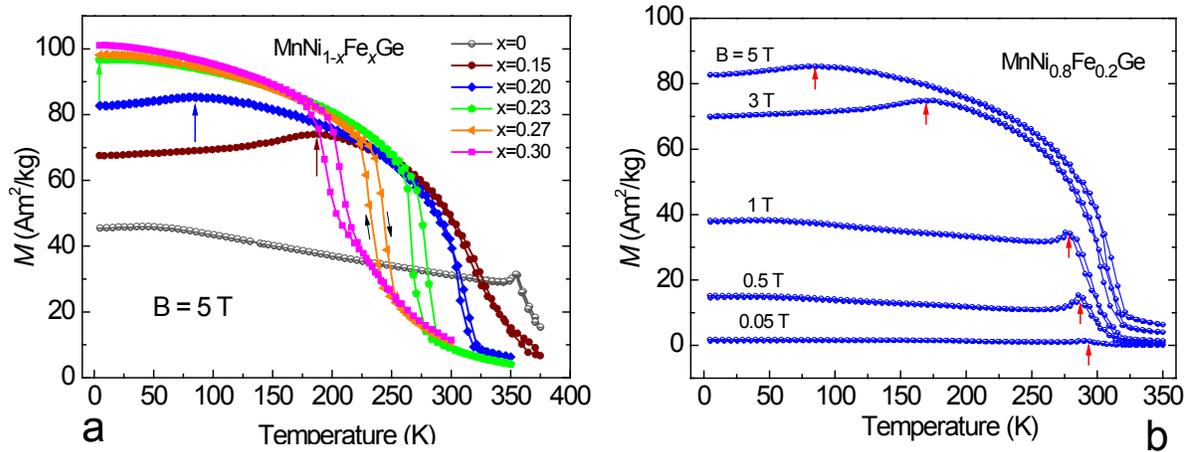

## Supplementary Figure S6:

**AFM-FM transition in Fe-substituted martensite of MnNi$_{1-x}$Fe$_x$Ge system. a**, Fe-content dependence of the critical temperature of the AFM-FM transition (marked by vertical arrows), measured in a field of 5 T. With increasing Fe content, the FM state is stable down to lower temperatures. **b**, Magnetic-field dependence of the critical temperature of the AFM-FM transition (marked by vertical arrows) for a typical sample with $x$ = 0.20. With increasing field, the FM state is stable down to low temperatures.

   Based on these data, a yellow-star line corresponding to the critical temperature ($T_{cr}^{5T}$) between FM and AFM states in a field of 5 T was indicated in phase diagram in Figure 3a in the main article. Upon cooling, the system in the range of 0.08 < $x$ < 0.23 returns back to an AFM state although a high field of 5 T is applied. This is due to the strong intrinsic AFM interactions in MnNiGe parent compound. Nevertheless, this AFM interaction has been effectively overcome by alloying FeNiGe into MnNiGe, as shown in Figure 3b in the main article.

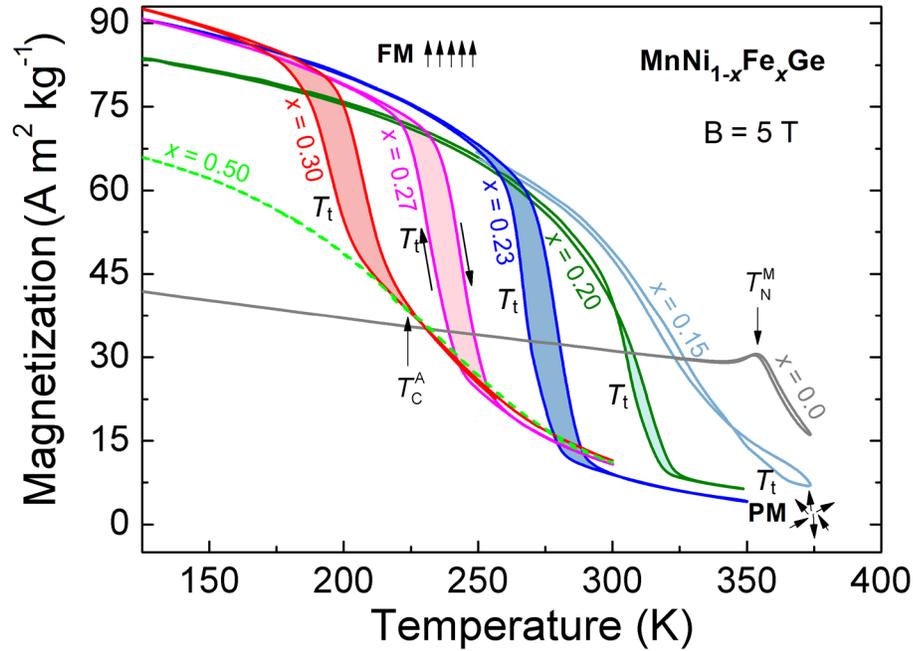

## Supplementary Figure S7:

**Magnetostructural coupling in the temperature window of MnNi$_{1-x}$Fe$_x$Ge system.**

The magnetization curves were measured in a high field of 5 T. $T_t$ indicates the martensitic-transition temperature. The gray curve is of the spiral AFM martensite of Fe-free stoichiometric MnNiGe. The $T_N^M$ in high field is at 350 K and the $T_t$ is at 460 K.

In accordance with the phase diagram, the martensitic transition temperature $T_t$ decreases with increasing Fe content. For the samples with x = 0.15 and 0.20, the martensitic transition temperature $T_t$s are just above the $T_C^M$ of martensites (see the phase diagram in main article), thus the martensitic and magnetic transitions are not completely coupled. For the sample with x = 0.30, the $T_C^A$ of austenite appears at about 225 K in the high field. Thus the martensitic transition occurs from FM austenite to FM martensite with a relatively small $\Delta M$. Inside the temperature window between

# SUPPLEMENTARY INFORMATION

$T_C^M$ and $T_C^A$, a sudden PM-FM jump of the magnetization based on the hysteretic martensitic transition is expectedly observed, which indicates that the introduction of Fe has led to a marked change of the magnetic interactions in the martensite and the AFM state of the martensite has thus been changed into a FM state.

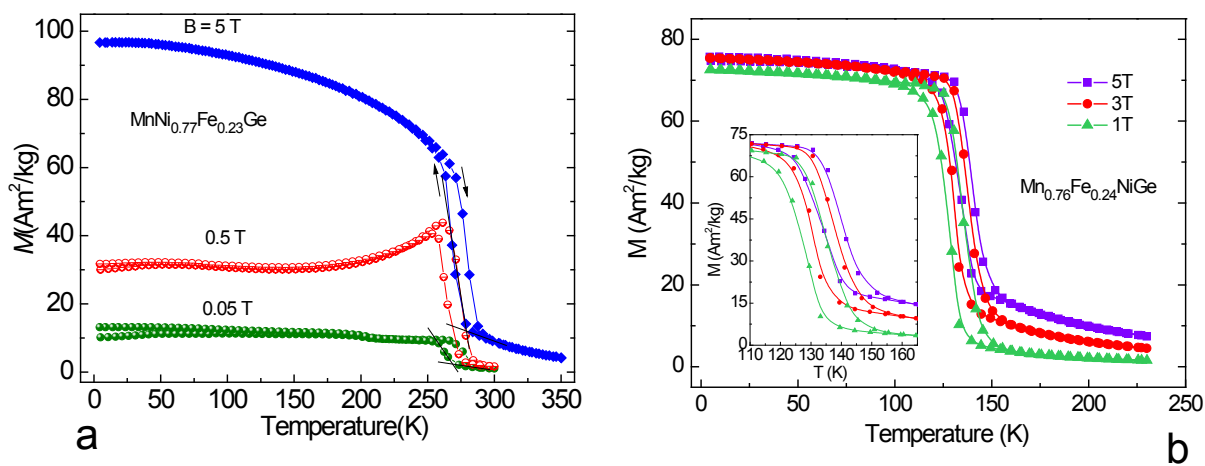

## Supplementary Figure S8:

**Thermomagnetization curves in different magnetic fields for the MnNi$_{0.77}$Fe$_{0.23}$Ge (a) and Mn$_{0.76}$Fe$_{0.24}$NiGe (b) samples.** With increasing applied field, the martensitic transition temperature was shifted toward high temperatures as the magnetic field supports the high-magnetization (martensite) phase. A field of 5 T shifts the starting temperature of martensitic transition by about 11 K for MnNi$_{0.77}$Fe$_{0.23}$Ge. The value for Mn$_{0.76}$Fe$_{0.24}$NiGe is 8 K in a field of 4 T. The magnetic-field-induced martensitic transition (MFIMT) behaviors can be seen in both MnNi$_{1-x}$Fe$_x$Ge and Mn$_{1-x}$Fe$_x$NiGe systems.

# SUPPLEMENTARY INFORMATION

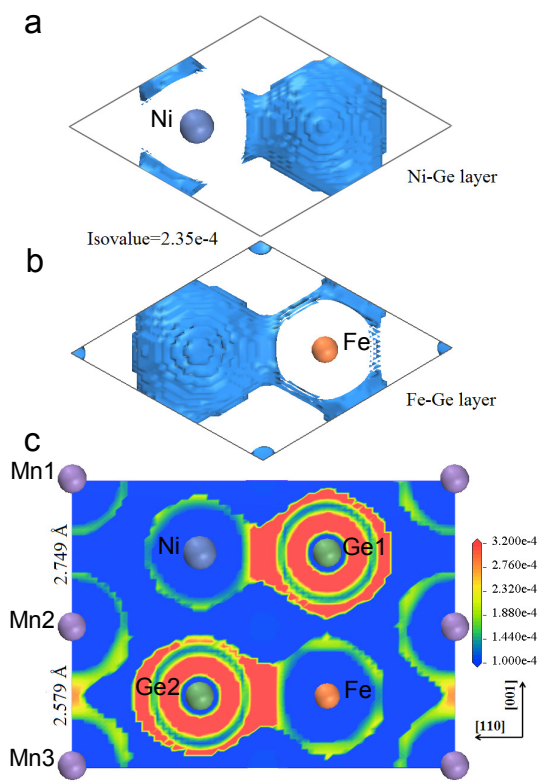

## Supplementary Figure S9:

**Valence-electron localization function (ELF) of MnNi$_{0.5}$Fe$_{0.5}$Ge austenite**. (Top view) ELF isosurface basins at an isovalue = 2.35×10$^{-4}$ of Ni-Ge (**a**) and Fe-Ge (**b**) layers. The Ge atoms are encapsulated by separate isosurfaces. **c**, ELF contour map in the (110) plane. The scale bar from blue to red corresponds to increasing electron localization.

The increasing stability of austenite phase may be greatly related to the strengthening of local chemical bonds when Fe atoms were introduced onto the Ni sites in MnNiGe. The valence-electron localization function (ELF)[5], as an indicator of the electron-pair distribution in terms of inter-atomic bonding, can provide the topological analysis of chemical bonds (see Supplementary Methods). Thus the ELF of MnNi$_{0.5}$Fe$_{0.5}$Ge has been calculated to probe the stability of austenite phase, as presented in Fig. S9. In this highly-ordered substituted structure, an alternating sequence of Fe-Ge and Ni-Ge layers is formed, as seen in Figs. S9a and S9b, which provides a convenient comparison between Fe-Ge and Ni-Ge bonding. From the topological analysis of ELF, it can be seen that the electron localization between nearest-neighbor Fe and Ge atoms (Ge1 in the Ni-Ge layers and Ge2 in the Fe-Ge layers) is strengthened with respect to that between nearest-neighbor Ni and Ge atoms. Furthermore, Fe-substitution leads to a considerable reduction of the interlayer spacing of (nearest-neighbor) Mn2 and Mn3 atoms, which is in agreement with the reduction of $c_h$ axis in XRD results (see Figure 2b in the main article). Thus, increased electron localization between the Mn atoms in interlayers is seen. These results indicate that alloying MnFeGe with MnNiGe leads to stronger covalence bonding between neighbor Fe and Ge atoms and between neighbor Mn and Mn atoms and therefore to an stabilization of the austenite phase. This stabilization is responsible for the decrease of the $T_t$ of the martensitic transition. Besides, the ELF calculations for alloying FeNiGe into MnNiGe give similar results.

# SUPPLEMENTARY INFORMATION

## Supplementary Methods

**ELF calculations.** *Ab initio* calculations were carried out using the pseudopotential method with plane-wave-basis set based on density-functional theory[6]. The electronic exchange correlation energy was treated under the local spin density approximation (LSDA)[7]. The Ge potential with electronic configuration of $3d^{10} 4s^2 4p^2$ was chosen, which also treats the Ge $3d$ outer-core electrons as valence electrons for higher accuracy. Plane-wave cut-off energy of 770 eV and 126 (13×13×12) k points in the irreducible Brillouin zone were used for a good convergence of the total energy. The absolute total-energy difference tolerance for the self-consistent field cycle was set at $5×10^{-7}$ eV/atom. The geometry optimizations for the atomic site occupancy in the cell were performed on the experimental lattice parameters using the Broyden-Fletcher-Goldfarb-Shanno (BFGS) minimization scheme[8]. The ELF[5] is introduced to represent the conditional probability of finding a second like-spin electron near the reference position. It is a local, relative measure of the Pauli repulsion effect on the kinetic energy density. Higher ELF values at the reference position show that the electrons are more localized than in a uniform electron gas of identical density. The topological analysis of ELF represents the organization of chemical bonds and, more particularly, the bond types.

# SUPPLEMENTARY INFORMATION

## Supplementary Table S1:

Curie temperature ($T_C^A$) of austenite in some near-stoichiometric MnNiGe systems from the references. The $T_C^A$ of stoichiometric MnNiGe is estimated at 205 K according to near-stoichiometric systems, which is below the Néel temperature of martensite ($T_N^M$) at 346 K. Thus, there is a large temperature interval of about 140 K between $T_C^A$ and $T_N^M$.

| Composition | $T_C^A$ (K) | $T_N^M$ (K) | Refs |
|---|---|---|---|
| $Mn_{1.05}Ni_{0.850}Ge$ | 205 | - | [9] |
| $Mn_{1.045}Ni_{0.855}Ge$ | 205 | - | [9] |
| $Mn_{1.091}Ni_{0.809}Ge$ | 198 | - | [10] |
| $Mn_{1.14}Ni_{0.76}Ge$ | 191 | - | [10] |
| $MnNiGe_{1.05}$ | 205 | - | [11] |
| MnNiGe | | 346 | [1] |
| | ≈205 | | Estimated |

## Supplementary Table S2:

Lattice constants and volumes of hexagonal and orthorhombic structures of $Mn_{0.84}Fe_{0.16}NiGe$ at the various temperatures. The subscripts "h" and "o" indicate the hexagonal and orthorhombic structures, respectively.

| Temp. (K) | $a_h$ (Å) | $c_h$ (Å) | $V_h$ (Å³) | $a_o$ (Å) | $b_o$ (Å) | $c_o$ (Å) | $c_o/\sqrt{3}$ (Å) | $V_o$ (Å³) | $V_o/2$ (Å³) | $(V_o/2 - V_h)/V_h$ (%) |
|---|---|---|---|---|---|---|---|---|---|---|
| 308 | 4.10181 | 5.37442 | 78.31 | - | - | - | - | - | - | - |
| 285 | 4.09863 | 5.36686 | 78.08 | - | - | - | - | - | - | - |
| 250 | 4.09796 | 5.36601 | 78.04 | - | - | - | - | - | - | - |
| 248 | 4.09618 | 5.36213 | 77.92 | - | - | - | - | - | - | - |
| 240 | 4.09571 | 5.35916 | 77.85 | 6.01605 | 3.74150 | 7.08077 | 4.08808 | 159.38 | 79.690 | 2.36 |
| 225 | 4.09555 | 5.35465 | 77.78 | 6.01157 | 3.74381 | 7.09696 | 4.09743 | 159.73 | 79.865 | 2.68 |
| 205 | 4.09646 | 5.35392 | 77.81 | 6.00970 | 3.74002 | 7.10205 | 4.10037 | 159.63 | 79.815 | 2.58 |
| 145 | - | - | - | 6.00921 | 3.73368 | 7.09737 | 4.09767 | 159.24 | 79.620 | - |
| 125 | - | - | - | 6.00985 | 3.73250 | 7.09803 | 4.09805 | 159.22 | 79.610 | - |
| 98 | - | - | - | 6.00835 | 3.72860 | 7.09437 | 4.09594 | 158.93 | 79.465 | - |

# SUPPLEMENTARY INFORMATION

## Supplementary Table S3:

Values of $T_t$ (cooling and heating), temperature-hysteresis of martensitic transition ($\Delta T$), Curie temperatures ($T_C^A$), Néel temperatures ($T_N^M$), martensitic magnetization ($M^M$) at 5 K, austenitic magnetization ($M^A$) at 5 K, magnetization difference ($\Delta M$) between austenite and martensite across the martensitic transition, critical field ($B_{cr}$) of metamagnetic transition in martensite phase and saturation field ($B_s$) of in martensite phase for $MnNi_{1-x}Fe_xGe$ and $Mn_{1-x}Fe_xNiGe$ samples. $T_g$ is the frozen temperature of spin-glass state of $Mn_{1-x}Fe_xNiGe$ samples ($x > 0.26$). $T_N^M$ were determined as the mean values of Néel temperatures on cooling and heating for every sample. $M^M$ and $M^A$ were measured at 5 K in a field of 5 T.

| Alloying case | x | $T_t$ (cooling) | $T_t$ (heating) | $\Delta T$ | $T_C^A$ ($T_g$) | $T_N^M$ | $M^M$ | $M^A$ | $\Delta M$ | $B_{cr}$ | $B_s$ |
|---|---|---|---|---|---|---|---|---|---|---|---|
| | | | | | K | | A m$^2$ kg$^{-1}$ | | | T | |
| MnNi$_{1-x}$Fe$_x$Ge | 0 | 470* | | - | - | 346* | 39* | - | - | - | >5* |
| | | 460 | 508 | 48 | - | 352 | 46 | - | - | 1.20 | >5 |
| | 0.10 | 400 | 416 | 16 | - | 324 | 52 | - | - | 0.90 | >5 |
| | 0.15 | 352 | 366 | 14 | - | 306 | 72 | - | - | 0.60 | >5 |
| | 0.20 | 300 | 312 | 12 | - | 291 | 87 | - | 46 | 0.45 | >5 |
| | 0.23 | 266 | 276 | 10 | - | - | 92 | - | 51 | 0.35 | >5 |
| | 0.27 | 230 | 240 | 10 | - | - | 95 | - | 49 | 0.20 | >5 |
| | 0.30 | 189 | 201 | 12 | 211 | - | 98 | - | 40 | 0.05 | 5.0 |
| | 0.33 | - | - | - | 213 | - | - | 77 | - | - | - |
| | 0.36 | - | - | - | 215 | - | - | 74 | - | - | - |
| | 0.40 | - | - | - | 218 | - | - | 72 | - | - | - |
| | 0.50 | - | - | - | 220 | - | - | 73 | - | - | - |
| Mn$_{1-x}$Fe$_x$NiGe | 0.03 | 429 | 472 | 43 | - | 349 | 52 | - | - | 1.13 | >5 |
| | 0.05 | 389 | 409 | 20 | - | 347 | 65 | - | - | 0.86 | >5 |
| | 0.08 | 336 | 357 | 21 | - | 341 | 75 | - | 36 | 0.57 | 5.0 |
| | 0.11 | 298 | 305 | 7 | - | - | 78 | - | 55 | 0.48 | 4.0 |
| | 0.13 | 277 | 285 | 8 | - | - | 80 | - | 57 | 0.45 | 3.0 |
| | 0.16 | 241 | 249 | 8 | - | - | 78 | - | 59 | 0.33 | 2.3 |
| | 0.18 | 207 | 214 | 7 | - | - | 79 | - | 59 | 0.24 | 1.8 |
| | 0.22 | 153 | 163 | 10 | - | - | 76 | - | 52 | 0.05 | 0.9 |
| | 0.24 | 127 | 135 | 8 | - | - | 75 | - | 54 | 0 | 0.5 |
| | 0.26 | 74 | 84 | 10 | - | - | - | 25.7 | - | - | - |
| | 0.28 | - | - | - | 70 | - | - | 23.0 | - | - | - |
| | 0.30 | - | - | - | 90 | - | - | 22.3 | - | - | - |
| | 0.35 | - | - | - | 90 | - | - | 17.6 | - | - | - |
| | 0.40 | - | - | - | 90 | - | - | 17.2 | - | - | - |
| | 0.50 | - | - | - | 85 | - | - | 18.1 | - | - | - |

* From Ref. S1. $M^M$ in Ref. S1 was measured at 90 K in a field of 5 T.

Supplementary references

# SUPPLEMENTARY INFORMATION

# SUPPLEMENTARY INFORMATION